\documentclass[preprint,12pt,double]{elsarticle}
\usepackage{ifpdf}
\usepackage{graphicx,natbib,amssymb,lineno}
\ifpdf
\usepackage[%
  pdftitle={Instructions for use of the document class
    elsart},%
  pdfauthor={Simon Pepping},%
  pdfsubject={The preprint document class elsart},%
  pdfkeywords={instructions for use, elsart, document class},%
  pdfstartview=FitH,%
  bookmarks=true,%
  bookmarksopen=true,%
  breaklinks=true,%
  colorlinks=true,%
  linkcolor=blue,anchorcolor=blue,%
  citecolor=blue,filecolor=blue,%
  menucolor=blue,pagecolor=blue,%
  urlcolor=blue]{hyperref}
\else
\usepackage[%
  breaklinks=true,%
  colorlinks=true,%
  linkcolor=blue,anchorcolor=blue,%
  citecolor=blue,filecolor=blue,%
  menucolor=blue,pagecolor=blue,%
  urlcolor=blue]{hyperref}
\fi

\makeatletter
\def\elsartstyle{%
    \def\normalsize{\@setfontsize\normalsize\@xiipt{14.5}}
    \def\small{\@setfontsize\small\@xipt{13.6}}
    \let\footnotesize=\small
    \def\large{\@setfontsize\large\@xivpt{18}}
    \def\Large{\@setfontsize\Large\@xviipt{22}}
    \skip\@mpfootins = 18\p@ \@plus 2\p@
    \normalsize
}
\@ifundefined{square}{}{}
\makeatother

\begin{document}
\newcommand{\cm}{cm$^{-1}$ }
\newcommand{\oC}{$^{\circ}$C }
\newcommand{\bcn}{Ba$_3$CaNb$_{2}$O$_9$ }
\newcommand{\complex}{A(B$'_{1/3}$B$''_{2/3}$)O$_3$ }
\newcommand{\tri}{$P\overline{3}m1$ }
\newcommand{\pero}{$Pm\overline{3}m$ }

\begin{frontmatter}

\title{Ordering and phonons in \bcn complex perovskite}

\author[phys_ufma]{Jo\~ao Elias Figueiredo Soares Rodrigues}
\author[rn]{Edvan Moreira}
\author[chem_ufma]{D\'ebora Morais Bezerra}
\author[chem_ufma]{Adeilton Pereira Maciel}
\author[phys_ufma,berk]{Carlos William de Araujo Paschoal \corref{corresp}}
\cortext[corresp]{Corresponding author: Phone: +55 98 3301 8222; Fax: +55 98 3301 8204; alternative e-mails: paschoal.william@gmail.com; paschoal.william@berkeley.edu}
\ead{paschoal@ufma.br}

\address[phys_ufma]{Departamento de F\'{\i}sica, CCET, Universidade Federal do Maranh\~ao, 65085-580, S\~ao Lu\'{\i}s - MA, Brazil}
\address[rn]{Universidade Federal Rural do Semi-\'Arido, UFERSA,    Campus Cara\'ubas, 59780-000, Cara\'ubas - RN, Brazil}
\address[chem_ufma]{Departamento de Qu\'{\i}mica, CCET, Universidade Federal do Maranh\~ao, 65085-580, S\~ao Lu\'{\i}s - MA, Brazil}
\address[phys_ufc]{Departamento de F\'{i}sica, Universidade Federal do Cear\'a, Campus do Pici, 60455-760, Fortaleza - CE, Brazil}
\address[berk]{Department of Materials Science and Engineering, University of California Berkeley, 94720-1760, Berkeley - CA, United States}

\begin{abstract}
In this work we performed a detailed investigation about ordering in \bcn perovskite. The sintering temperature and time were changed to obtain samples with different ordering. The order parameters were probed by Raman spectroscopy  based on a partial disordered model. To use the partial disordered model correctly we performed {\it ab initio} calculations in \bcn to assign the optical phonons. The results showed that sintering temperature improves order while sintering time is not so efficient to promote order.
\end{abstract}

\end{frontmatter}

\section{Introduction}

Complex triple perovskites with the general formula A$_3$B$'$B$''_2$O$_9$, where A is an alkaline earth metal, B$'$ and B$''$ are ions with valencies $2+$ and $5+$, have recently attracted much interest as high-temperature proton conductors \cite{Ruiz-Trejo2005,Nowick,Valdez-Ramirez2012,Bhella2011,OIKAWA,Trinh2010,Mani2007a} and as dielectric resonators \cite{Bhalla2000a,Dias2003g,Sagala1993b,Dias2003,Lufaso2005,Yue2004,Chen2006}. Particularly, the nonstoichiometric series Ba$_3$Ca$_{1+x}$Nb$_{2−x}$O$_{(9-3x)/2}$ has been extensively investigated due to its high proton conductivity and chemical stability \cite{Ruiz-Trejo2005,Bhella2011,OIKAWA,Trinh2010,Mani2007a}. In such oxides, the formation of oxygen vacancies is achieved by increasing the $B'/B''$ ratio. The vacancies can be filled with OH$^−$ species upon annealing samples in water vapour at high temperatures and even OH$^-$ concentration is small, the ionic conductivity can be dominated by proton transport \cite{Kreuer1996}, enabling it for application in fuel cells.

Despite oxygen vacancies can be highly mobile in simple ABO$_3$ perovskites,\cite{Boukamp2003} in complex perovskites as \bcn the vacancy mobility tends to be strongly influenced by the $B'/B''$ ratio, valencies and ordering. Substitutions in A and B sites of ABX$_3$ perovskites are responsible for improving a large part of the applications of these materials, bringing new physical and chemical properties.\cite{Bhalla2000a} However, the substitution by more than one ion, mainly in B site, can lead to structures which are disordered or ordered, according the new ions distribution into the structure. This substitution leads to new crystalline structures when there is a ordered distribution. B-site cation-ordered triple perovskites whose chemical formula is A$_3$B$'$B$''_2$O$_9$ are a particular family obtained by B-site substitution .

From the structural viewpoint, in the triple B-site ordered substitution (sometimes called 1:2 order), two main structures can be formed. In the First, an ordered B$'$/B$''$ substitution into B site can lead to a trigonal structure which belongs to the space group $P\overline{3}m1$. In this structure the cations  B$'$ and B$''$ are alternately distributed in $\{111\}$ planes in the form $\cdot \cdot \cdot B'B''B''B'B''B''B' \cdot \cdot \cdot$. Another possible structure for this composition is the so-called hexagonal $6H$ structure that belongs to the $P6_3/mmc$ space group, in which the B$'$O$_6$ and B$''$O$_6$ octahedra share edges.\cite{Mani2007a,Lufaso2005}

However, partial disordered B$'$/B$''$ distribution in both structure can occur, with part of the B$'$ cations occupying the B$''$ cation sites and vice versa. Partial disorder in perovskites plays an essential role because they strongly influence their physical properties. For example, order influences phonons and dielectric constant, implying in consequences for applications of 1:2 perovskites in wireless communication system.\cite{Truong2007a,Lee2007a,Davies2005,Akbas2005,Levin2001a,KimI.T.Kim1997,Dias2003,Dias2003g,Rodrigues} Disorder is particularly interesting in \bcn perovskites because it favors a higher conductivity in these protonic conductors based on nonstoichiometry Ba compounds, which always have higher conductivity and lower activation energy than the corresponding Sr compounds.\cite{Nowick} Due to the importance of ordering in \bcn (BCN) ceramics, mainly because its conductivity properties, in this work we proposed a systematically investigation of the order achievement in BCN complex perovskite.

\section{Experimental Procedures}
BCN samples were synthesized by polymeric precursor method using barium nitrate (Ba(NO$_3$)$_2$, Sigma Aldrich), calcium citrate tetrahydrate (Ca$_3$(C$_6$H$_5$O$_7$)$_2 \cdot $4H$_2$O, Ecibra) and ammonium complex of niobium (NH$_4$(NbO(C$_2$O$_4$)$_2$(H$_2$O)$_2$) $ \cdot $3H$_2$O, CBMM) as sources of metals \cite{pechini}. Barium polymeric precursor was obtained by mixing aqueous solutions of barium nitrate  and citric acid (C$_6$H$_8$O$_7 \cdot$ H$_2$O, Proqu\'{\i}mico) in a molar ratio of 1:3 metal-citric acid, keeping stirring between 60 and 70 \oC. Finally, ethylene glycol (HOCH$_2$CH$_2$OH, Merck) was added to metal/Citric aqueous solution in a mass ratio of 1:1 in relation to citric acid. Calcium polymeric precursor was obtained following the same procedure for barium. To obtain niobium precursor we firstly precipitated niobium oxi-hydroxide stirring an aqueous solution of ammonium complex of niobium until pH of 9 in a thermal bath at 0 \oC. Niobium hydroxide (Nb(OH)$_5$) was separated from oxalate ions using distilled water at 40-50 \oC under vacuum filtering. Finally, citric acid and ethylene glycol were added to the Niobium hydroxide aqueous solution. The pH of the polymeric precursors was kept at the same values to avoid precipitations when mixing them. To determine the necessary precursor weight to achieve the correct metal stoichiometry to form BCN perovskite we used gravimetric analysis at 900 \oC for 1 h. We mixed the three precursors and heated the mixture between 80 and 90 \oC to form the polyester resin, which had high viscosity and glassy. The resin was annealed  at 400 \oC for 2h. This heat treatment converted the resin in a black porous powder. This powder was lightly grounded using an agate mortar. After it was calcined at 1300 \oC for 2h to obtain BCN sample. This sample is our reference to investigate the ordering process and it will be referenced as {\it start sample} along this work. From this reference sample, several pellets were sintered at different temperatures and time of sintering to investigate the ordering phenomena.

The crystalline structure of the samples were probed by powder X-ray diffraction (XRD - Bruker D8 Advance). We performed a continuous scanning mode using Cu-K$_{\alpha 1}$ radiation (40 kV, 40 mA) over a $2\theta$ range between $10^\circ$ and $100 ^\circ$ (0.02$^ \circ$/step with 8 seconds/step). The powder XRD pattern was compared with data from ICSD (Inorganic Crystal Structure Database, FIZ Karlsruhe and NIST) International diffraction database (ICSD\#162758). \cite{Deng2009}  The structure was refined  using the DBWS9807 free software. \cite{Rietveld:a05418,Rietveld:a07067}

The Raman spectra of the samples were acquired at room temperature in an iHR550 Horiba scientific spectrometer coupled to an Olympus  microscope model BX-41. A He-Ne laser (632.8 nm, 10 mW) was use to excite the spectra that were collected in an air-cooled Synapse CCD detector. The spectral resolution was kept lower than 2 \cm using an 1800 grooves/mm grating in the spectrometer. All spectra were acquired in a backscattered configuration.

\section{Computational method details}

The first-principle calculations were performed through the Cambridge Serial Total Energy Package (CASTEP) software package \cite{Segall2002}. Density functional theory (DFT) \cite{Kohn1965,Hohenberg1964} was chosen to model the material considering generalized gradient approximation (GGA) as exchange-correlation functional. This approximation was improved by the Perdew-Burke-Ernzerhof (PBE) parameterization \cite{Perdew1996}. The PBE functional leads to results close to those obtained  using PW91 functional \cite{Perdew1992} within TS method \cite{Tkatchenko2009} for the dispersion correction scheme (DFT+D) to describe van der Waals interactions. We also adopted pseudopotentials to replace the core electrons in each atomic species. Specifically, norm-conserved pseudopotentials \cite{Lin1993} were used. These pseudopotentials were generated using the OPIUM code \cite{Rappe1990}, following the same scheme of previous works \cite{Moreira2012a,Moreira2011,Moreira2012}. The electronic valence configurations for each atomic species were: Ba$-5s^25p^66s^2$, Ca$-3s^23p^64s^2$, Nb$-4d^45s^1$, and O$-2s^22p^4$. A Monkhorst-Pack \cite{Monkhorst1976} $3\times3\times2$ sampling was used to evaluate all integrals in the reciprocal space. This sampling is enough to give a well converged electronic structure.

We employed BFGS minimizer \cite{Pfrommer1997} to optimize the unit cell. In this minimization algorithm, a starting Hessian is recursively updated. For each self-consistent field step, the electronic minimization parameters were: Total energy/atom convergence tolerance of $0.5\times10^{-6}$ eV, a maximum energy eigenvalues threshold of $0.1442\times10^{-6}$ eV, and a convergence window of 3 cycles. A plane-wave basis set was adopted to represent the Kohn-Sham orbitals, with cutoff energy chosen of 880 eV. This value was obtained after convergence studies. The quality of this basis set was kept fixed as the unit cell volume varies during geometry optimization.

The vibrational properties of BCN were calculated performing density functional perturbation theory (DFPT) calculations or linear response formalism \cite{Baroni2001}. The linear response provides an analytical way of computing the second derivative of the total energy with respect to a given perturbation. Depending on the nature of this perturbation, a number of properties can be calculated. A perturbation in ionic positions gives the dynamical matrix and phonons; in the presence of magnetic field it gives the NMR response; in unit cell vectors changes it gives the elastic constants; in an electric field presence it gives the dielectric response, etc. The infrared absorption intensities are described in terms of a dynamical matrix and Born effective charges (also known as atomic polarizability tensors, ATP) \cite{Baroni2001} and can be obtained by calculating the phonons at the $\Gamma$ point $(\vec{k}=0)$. The structure used for vibrational calculations was the fully ordered trigonal BCN structure optimized through the GGA-PBE approximation. The geometry optimization criteria were more rigorous than the criteria used for LDA-DFT calculations, for example. The convergence thresholds used were: Total energy convergence tolerance smaller than $5\times10^{-6}$ eV/atom, maximum ionic force smaller than $10^{-2}$ eV/\AA, maximum ionic displacement tolerance of $5\times10^{-4}$ \AA, and maximum stress component smaller than $2\times10^{-2}$ GPa. For the self-consistent field calculations, the convergence criteria considered a total energy per atom variation smaller than $5 \times 10^{-7}$ eV, and electronic energy eigenvalue variation smaller than $0.1442 \times 10^{-6}$ eV.

\section{Results and discussion}
\begin{figure}
  \centering
  \includegraphics[width=10cm]{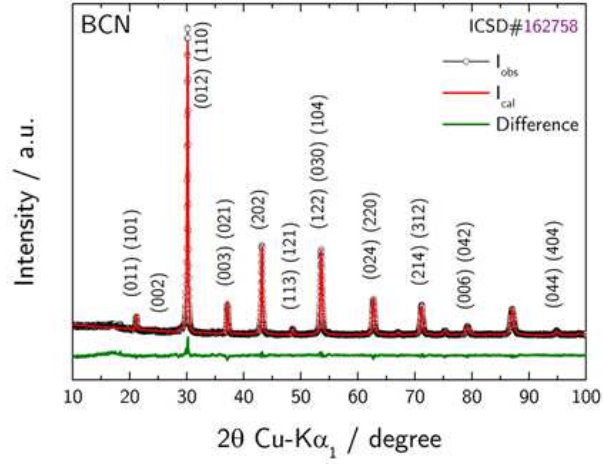}\\
  \caption{Powder X-ray diffraction pattern of BCN ceramic sintered for 1300 \oC/2h. The solid line is the fitting using the Rietveld method and difference between the experimental and calculated patterns. The indices of planes are labeled in picture.}\label{drx}
\end{figure}

Figure \ref{drx} shows the powder X-ray diffraction pattern obtained for the {\it start sample}, which was sintered at 1300\oC for 2h. The reflection planes were indexed according to the trigonal structure which belongs to the space group $P\overline{3}m1$. The  refinement parameters were summarized in Table \ref{data_drx}. In this structure the cations Ca$^{2+}$ and Nb$^{5+}$ are alternately distributed in $\{111\}$ planes in the form $\cdot \cdot \cdot Ca-Nb-Nb-Ca-Nb-Nb \cdot \cdot \cdot$. This distribution produces characteristic reflections of the trigonal superstructure.\cite{GalassoF.BarranteJ.R.Katz1961} When Ca and Nb are randomly distributed, and consequently the structure is disordered, the powder pattern does not exhibit such reflections, inducing changes in the plane indexing. In this case, a cubic structure that belongs to the space group $Pm\overline{3}m$ is usually observed.
\begin{table}
  \centering
  \begin{tabular}{ll}
         \hline \hline
Space group; Z					     			  & $P\overline{3}m1$ (N$^o$ 164); 3   \\
Lattice parameters, \AA					          & a=5.8921(8), c=7.2358(8)           \\
Temperature, K								      & 298(1)                             \\
Density (calculated), g/cm$^3$		              & 5.965                              \\
Volume, \AA$^3$									  & 217.56                             \\
$\lambda$, \AA									  & 1.54056                            \\
Profile function 							      & Pseudo-Voigt                       \\
Cagglioti parameters (U, V, W)	                  & 0.2056(2), -0.0496(8), 0.0549(6)   \\
Background function							      &	5th order polynomial in 2$\theta$         \\
Rp, Rwp, Rexp, \%								  & 7.63, 10.05, 5.90                  \\
\hline \hline
   \end{tabular}
  \caption{Data collection and Refinement Details for BCN sample sintered for 1300\oC for 2h.}\label{data_drx}
\end{table}

\begin{figure}
  \centering
  \includegraphics[width=10cm]{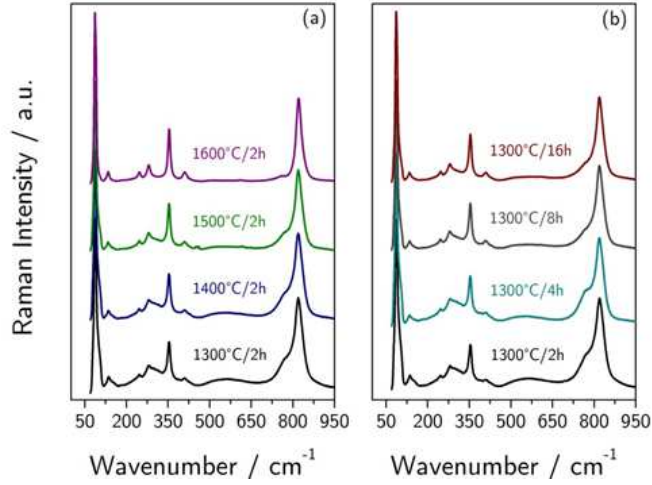}\\
  \caption{Raman spectra of the BCN ceramics sintered for several values of sintering temperature and sintering time.}\label{raman_sinter}
\end{figure}

To investigate the formation of ordered BCN structure we performed Raman scattering measurements in BCN samples sintered for several values of sintering temperature and sintering time. The observed Raman spectra are shown in Figure \ref{raman_sinter}.
To classify and identify the phonons observed we used a group theoretical analysis. Thus, we determined the symmetry and number of phonons in the full ordered trigonal structure based on the irreducible representation of the group factor $m\overline{3}m$. \cite{Rousseau1981a} In this structure the primitive unit cell has 15 atoms, where one barium ion is at the $1a$ site and two others are at the $2d$ site, the calcium ion is at $1b$ site and the two niobium ions are at $2d$ site, six oxygen ions are at $6i$ site and three are at the $3e$ site.\cite{Deng2009,Deng2009a} Therefore, based on this ion occupation, the expected phonons in BCN are shown in Table \ref{modes}. However, nine Raman-active modes $(4A_{1g} \oplus 5E_{g } )$ are expected in BCN at room temperature.
\begin{table}
  \centering
  \begin{tabular}{cccl}
\hline \hline
{\bf Ion}	& {\bf Site}  &  {\bf Symmetry}	  &  {\bf Distribution of modes} \\ \hline
Ba1	& 1a	  &  -3m	      &  $ A_{2u} \oplus  E_{u }  $                            \\
Ba2	& 2d	  &  3m		      &  $ A_{1g} \oplus  A_{2u} \oplus   E_{g} \oplus  E_{u }  $                  \\
Ca1	& 1b	  &  -3m	      &  $ A_{2u} \oplus  E_{u } \oplus $                         \\
Nb1	& 2d	  &  3m		      &  $ A_{1g} \oplus  A_{2u} \oplus   E_{g} \oplus  E_{u } $                   \\
O1	& 3e	  &  2/m	      &  $ A_{1u} \oplus 2A_{2u} \oplus  3E_{u}     $                   \\
O2	& 6i	  &  m		      &  $2A_{1g} \oplus  A_{1u} \oplus  A2_{g} \oplus 6A_{2u} \oplus 3_{Eg} \oplus 3 E_u $ \\ \hline
		&       &  $\Gamma$		      &  $4A_{1g} \oplus 2A_{1u} \oplus  A2_{g}  \oplus8A_{2u} \oplus 5_{Eg} \oplus 10E_u $  \\
		&       &  $\Gamma^{Acoustic}$	&  $ A_{2u} \oplus  E_{u } $                              \\
		&       &  $\Gamma^{Raman}$			&  $4A_{1g} \oplus 5E_{g } $                            \\
		&       &  $\Gamma^{IR	}$			&  $7A_{2u} \oplus 9E_{u } $                            \\
		&       &  $\Gamma^{Silent	}$	&  $2A_{1u} \oplus  A_{2g} $                             \\
\hline \hline
  \end{tabular}
  \caption{Distribution of modes in the perovskite crystalline structure belonging to the trigonal space group $P\overline{3}m1$.}\label{modes}
\end{table}

In another way, occupational disorder can occur in two different ways: It implies in extra optical-active modes due to the different sites occupied by the ions (two-phonon behavior),\cite{Dias2001} or it implies in local symmetry lowering (one phonon behavior),\cite{Ayala2002a} changing the Raman and infrared spectrum. In this work we assumed the first case, that is the most common in 1:2 perovskites.\cite{Moreira2001,Dias2001} Thus, to index the BCN crystalline structure according to the trigonal structure implies in an order-disorder model in which the trigonal perovskite structure is not fully ordered, although it still has a trigonal unit cell.\cite{Chia2003,Moreira2001,Chen2006,Surendran2005} In this case, there are part of calcium ion occupying the site $2d$ and part of niobium ion at the site $1b$.\cite{SinyI.G.TaoR.KatiyarR.S.GuoR.Bhalla1998,Siny1998} Thus, in this order-disorder model, the group theoretical analysis can be used to predict the extra optical-active phonons based on extra site occupation by Ca and Nb ions, as showed in Table \ref{disorder}. Therefore, considering the disorder, there are now eleven Raman-active phonons in the partially ordered BCN.
\begin{table}
  \centering
    \begin{tabular}{cccl}
      \hline \hline
{\bf Ion}	& {\bf Site}  &  {\bf Symmetry}	  &  {\bf Distribution of modes} \\ \hline
Ca2	      &     2d	    & 3m	&   $A_{1g} \oplus A_{2u} \oplus E_g \oplus E_u $\\
Nb2	      &     1b	    & -3m	&   $A_{2u} \oplus E_u $ \\
      \hline  \hline
    \end{tabular}
  \caption{Distribution of modes for additional sites in the order-disorder model for the 1:2 perovskite compounds partially ordered.}\label{disorder}
\end{table}

From the Figure \ref{raman_sinter} (See bottom spectra in Figures \ref{raman_sinter}(a) and \ref{raman_sinter}(b)) we observed fourteen vibrational modes in the start sample, confirming a partially ordered 1:2 perovskite structure \cite{Moreira2001,Dias2001} (see Figure \ref{raman_sinter}b). Clearly, BCN Raman spectra change significatively when the sintering temperature and time are increased with both playing an essential role in ordering process.

The most ordered sample is achieved when sintered at 1600 \oC for 2h, whose spectral deconvolution is shown in Figure \ref{deco} and summarized in Table \ref{comparison}.

\begin{figure}
  \centering
  \includegraphics[width=10cm]{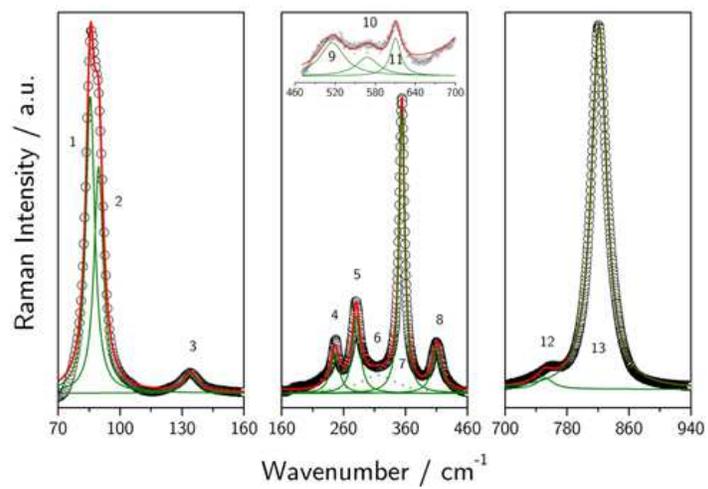}\\
  \caption{Deconvolution of the Raman spectra of the BCN pellet sintered at 1600\oC/2h. Collected data (o) and calculated Lorentzian (-) curve by fitting.}\label{deco}
\end{figure}
\begin{table}[h]
  \centering
  \caption{Raman- and IR-active modes of trigonal BCN structure with respective attribution. A comparison between experimental (EXP.) and calculated (CAL.) data is shown to Raman modes. $^a$FLB: Floating Base Line; $^b$DAM: Defect activated IR-active modes.}\label{comparison}
\begin{tabular}{ccccccc}
  \hline   \hline
 \multicolumn{4}{c}{Raman-active modes} & & \multicolumn{2}{c}{Infrared-active modes} \\ \cline{1-4}\cline{6-7}
Peak & Center (EXP.)/\cm &	Center (CAL.)/\cm &	Attrib.	& &	Center (CAL.)/\cm &	Attrib. \\ \cline{1-4}\cline{6-7}
1 &	85.6	&59.4	&  $E_{g}$		  &	& 58.9	 &  $A_{2u}$  \\
2 &	89.6	&60.4	&  $A_{1g}$	  &	& 77.5	 &  $E_{u}$   \\
3 &	134.2	&62.0	&  $E_{g}$		  &	& 85.9	 &  $A_{2u}$  \\
4 &	245.6	&241.2	&  $A_{1g}$	&	& 101.4	 &  $E_{u}$   \\
5 &	280.1	&250.7	&  $E_{g}$		&	& 136.9	 &  $E_{u}$   \\
6 &	315.4	&-			&  FLB$^a$	&	& 179.6	 &  $E_{u}$   \\
7 &	353.9	&305.8	&  $E_{g}$		&	& 191.8	 &  $E_{u}$   \\
8 &	410.5	&356.8	&  $A_{1g}$	&	& 223.1	 &  $A_{2u}$  \\
9 &	515.9	&-			&  DAM$^b$	&	& 283.7	 &  $E_{u}$   \\
10&	568.4	&-			&  $E_{g}$		&	& 291.2	 &  $A_{2u}$  \\
11&	610.5	&658.9	&  			&	& 352.4	 &  $E_{u}$   \\
12&	751.5	&-			&  $A_{1g}$	&	& 374.9	 &  $A_{2u}$  \\
13&	821.2	&827.7	&  			&	& 503.7	 &  $E_{u}$   \\
	&				&584.5	&  $A_{2u}$  & &         &        \\
	&				&663.1	&  $E_{u}$   & &         &        \\
	&				&807.8	&  $A_{2u}$  & &         &        \\ \hline \hline
\end{tabular}
\end{table}

To identify the Raman-active phonons in BCN and check the partial disordered model, we performed {\it ab initio} calculations of the vibrational properties of the full ordered BCN trigonal structure. The good reliability of the method employed can be seen comparing the experimental\cite{Deng2009} and calculated structural data given in Table \ref{calc}. The calculated phonons are summarized in Table \ref{comparison} together with the experimental phonons. The infrared-active phonons are also shown for completeness.
\begin{table}
  \centering
  \begin{tabular}{cccc}
     \hline \hline
     a / \AA	     &  c / \AA	&  $\gamma$	&V / \AA$^3$   \\
GGA	    6.0279 & 7.4631	&  120$^o$	  &  234.849     \\
EXP.	5.9037 & 7.2636	&  120$^o$	  &  219.246     \\
     \hline \hline
   \end{tabular}
  \caption{Lattice parameters for BCN compound calculated within GGA-PBE calculation using the norm-conserving pseudopotential. Experimental data (EXP.) for BCN ceramic are also presented.\cite{Deng2009}}\label{calc}
\end{table}

The occurrence of the peaks n$^\circ$ 10 and 12 is correlated to the partial ordering adopted by the BCN confirming the disorder model adopted. This behavior can be explained by the Nb-O bond length decreases as consequence the Nb$^{5+}$ substitution by the Ca$^{2+}$, resulting in a displacement of the A$_{1g}$ (821 \cm) mode to lower wavenumber.\cite{Deng2009a} Also, a simple harmonic model to these vibrations, based on the charge/mass relation $\frac{[q_{Ca}/m_{Ca}]^{1/2}}{[q_{Nb}/m_{Nb}]^{1/2}}$ give us the value 0.96, showing that the $A_{1g}$ mode due to the extra-site Ca$^{2+}$ occupation should be observed 0.96 times that $A_{1g}$ band associate to the Nb$^{5+}$ correct occupation. \cite{Moreira2001}  The calculated ratios, between the peaks n$^\circ$ 12 and 13 (0.92) and n$^\circ$ 10 and 11 (0.93) (see Table \ref{comparison}), show this assumption is correct.

The evaluation of the ordering kinetics of the BCN structure was followed according the evolution of the $A_{1g}$ modes under the sintering temperature and time changes. A detailed spectral deconvolution of these modes are shown in Figure \ref{deco}.
\begin{figure}
  \centering
  \includegraphics[width=10cm]{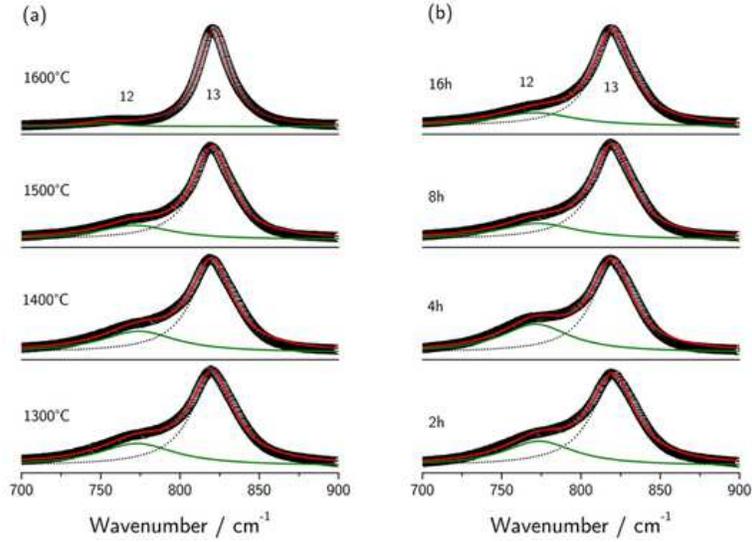}\\
  \caption{Deconvolution of the Raman spectra of the BCN ceramics in the spectral range 700-900 \cm for several values of sintering temperature (a) and sintering time (b). Collected data (o) and calculated Lorentzian (-) curve by fitting.}\label{deco}
\end{figure}

\begin{figure}
  \centering
  \includegraphics[width=10cm]{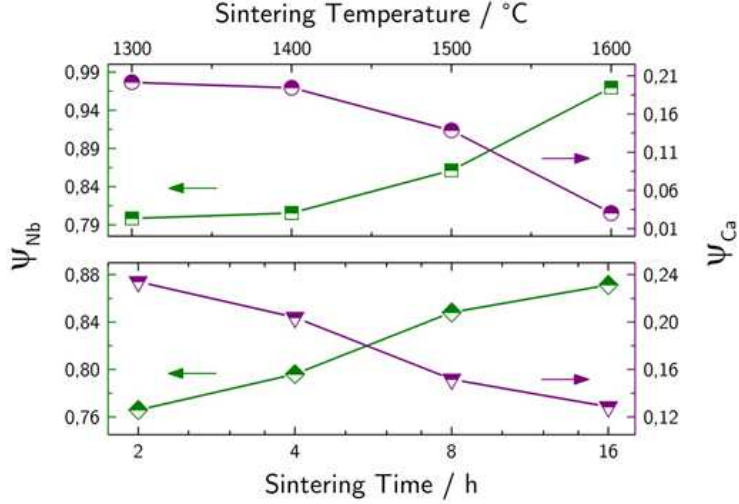}\\
  \caption{Behavior of the ratio $\Psi_{Ca,Nb}$ in function of the sintering temperature and of the sintering time for BCN ceramics.}\label{sinter}
\end{figure}

Clearly, the intensities of the peak n$^\circ$ 10 and 12 are proportional to the percentage of the extra-site Ca$^{2+}$ and Nb$^{5+}$ ions. Using the modes n$^\circ$ 12 and 13 we can estimate the ordering degree setting a ratio designated $\Psi_{Ca,Nb}$ expressed by
\begin{equation}\label{order_degree}
  \Psi_{Ca,Nb}=\frac{I_{12,13}}{I_{12}+I_{13}}
\end{equation}
When $\Psi_{Nb}$ is equal to one, all Nb ions are in correct site. In this case $\Psi_{Ca}$ is null. Thus, the fully ordered BCN structure occurs when $\Psi_{Ca}=I_{12}=0$ and $\Psi_{Nb}=I_{13}=1$. Figure \ref{sinter} shows the evolution of the ratio $\Psi_{Ca,Nb}$ under the sintering parameter changes. The most ordered sample is that sintered at 1600 \oC for 2h for which $\Psi_{Ca} (0.03)$ and $\Psi_{Nb} (0.97)$. We observe the sintering time is not so efficient to order BCN, once that the calculated ratios to BCN ceramic sintered at  1300 \oC for 16h were $\Psi_{Ca} (0.13)$ and $\Psi_{Nb} (0.87)$.

\section{Conclusions}

Partially ordered BCN ceramics was obtained by polymeric precursor method and the ordering in these samples were investigated by Raman spectroscopy under sintering temperature and sintering time changes. The evolution of the order was evaluated with basis on a partially ordered trigonal structure. An ab initio calculation permit us to assign the phonons and to monitor the order observing the changes in the behavior of the $A_{1g}$ Raman-active mode near to 821 \cm. The most ordered sample was obtained at a sintering temperature of 1600 \oC at 2 h.

\section*{Acknowledgement}
The authors are grateful to the Brazilian funding agencies CAPES, CNPq, INCT NANOBIOSIMES and FAPEMA.


\end{document}